\documentclass[twocolumn,showpacs,preprintnumbers,amsmath,amssymb,prb,superscriptaddress]{revtex4}

\usepackage{graphicx}
\usepackage{dcolumn}
\usepackage{bm}

\begin{document}

\title{Interaction Correction to the Longitudinal Conductivity and Hall Resistivity in High Quality Two-Dimensional GaAs Electron and Hole Systems}

\author{C. E. Yasin}\author{T. L. Sobey}\author{A. P. Micolich}\author{A. R. Hamilton}\author{M. Y. Simmons}
\affiliation{School of Physics, University of New South Wales, Sydney 2052, Australia}

\author{L. N. Pfeiffer}\author{K. W. West}
\affiliation{Bell Laboratories, Lucent Technologies, Murray Hills NJ 07974, USA}

\author{E. H. Linfield}\author{M. Pepper}\author{D. A. Ritchie}
\affiliation{Cavendish Laboratory, University of Cambridge, Cambridge, CB3 0HE, United Kingdom}

\date{\today}

\begin{abstract}
We present a systematic study of the corrections to both the longitudinal conductivity and Hall resistivity due to
electron-electron interactions in high quality GaAs systems using the recent theory of Zala \emph{et al.} [Phys. Rev. B
\textbf{64}, 214204 (2001)]. We demonstrate that the interaction corrections to the longitudinal conductivity and Hall resistivity
predicted by the theory are consistent with each other. This suggests that the anomalous metallic drop in resistivity at $B$=0 is
due to interaction effects and supports the theory of Zala \emph{et al}.
\end{abstract}


\maketitle

Despite much theoretical and experimental work, to date the origin of the anomalous metallic behavior in high quality 2D systems at
zero magnetic field $B$ remains a subject of controversy \cite{Review}. The earliest results of Finkelstein \cite{Finkelstein} and
Castellani \emph{et al.} \cite{Castellani} suggested that in the presence of weak disorder and strong interactions it is possible
to have a metallic ground state (where the conductivity $\sigma_{xx}$ assumes a finite value at $T$=0) in 2D, even at $B$=0. It is
well known that in the diffusive limit $k_BT\tau/\hbar \ll$ 1, electron-electron interactions give rise to a logarithmic correction
to the conductivity. \cite{Altshuler} In the opposite ballistic limit $k_BT\tau/\hbar \gg$ 1, Stern, \cite{Stern} Gold $\&$
Dolgopolov, \cite{Gold} and Das Sarma $\&$ Hwang \cite{DasSarma} have also shown that $\sigma_{xx}$ can exhibit a metallic-like
temperature dependence that is linear in temperature, and due to temperature-dependent screening of impurity and interface
roughness scattering. While these previous theories were formulated in opposite limits of $k_BT\tau/\hbar$, most existing
experimental data displaying $B$=0 metallic behavior falls in the intermediate region between these limits.

Recently, quantum conductivity corrections have received renewed interest following the development of a theory by Zala \emph{et
al.} \cite{Zala} that is valid for the entire range of $k_BT\tau/\hbar$. This theory shows that the previous results in the two
opposite limits are due to the same physical process -- elastic scattering of electrons by the self-consistent potential created by
all the other electrons (i.e., scattering from the screened impurity potential). As the temperature is reduced, Ref.
\onlinecite{Zala} predicts a correction to the conductivity that is either localizing or delocalizing depending on the value of the
Fermi liquid parameter $F^\sigma_0$. The parameter $F^\sigma_0$ is a measure of the interaction strength in the triplet channel and
is related to the spin susceptibility $\chi \propto 1/(1+F^\sigma_0)$ \cite{Zala}. Several recent experimental studies
\cite{Proskuryakov,Pudalov,Coleridge,Noh} following Ref. \onlinecite{Zala} have shown that the $B$=0 metallic behavior in a variety
of material systems is consistent with the theory of Zala \emph{et al.}

In addition to producing a correction to the longitudinal conductivity $\sigma_{xx}$, interaction effects also lead to a correction
to the Hall resistivity $\rho_{xy}$. \cite{Altshuler,Zala_Hall} Whilst many of these authors have examined the $\sigma_{xx}$
interaction correction, both at $B$=0 and in parallel magnetic fields, \cite{Proskuryakov,Pudalov,Coleridge,Noh,Vitkalov} studies
of the $\rho_{xy}$ correction are less common. \cite{Coleridge,Simmons} In particular there has been no experimental confirmation
that the observed correction to $\rho_{xy}$ is consistent with the correction to $\sigma_{xx}$ or with the theoretical predictions
of Ref. \onlinecite{Zala_Hall}. Since the magnitude of the corrections to $\sigma_{xx}$ and $\rho_{xy}$ both depend on
$F^\sigma_0$, then if the $B$=0 metallic behavior is due to electron-electron interactions, the two values of $F^\sigma_0$
extracted separately from measurements of $\sigma_{xx}$ and $\rho_{xy}$ should be consistent.

In this paper we report a systematic study of the interaction corrections to both the longitudinal conductivity $\sigma_{xx}$ and
Hall resistivity $\rho_{xy}$, and extracted values for the Fermi liquid parameter, $F^\sigma_0$. We show that the values of
$F^\sigma_0$ obtained from these two independent measurements are consistent. Our results support the theory of Zala \emph{et al.}
and suggest that the $B$=0 metallic behavior is due to interaction effects.

The experiments were performed on both n- and p-GaAs samples. The measured mobilities were 8.0 $\times 10^5$ cm$^2$V$^{-1}$s$^{-1}$
at a carrier density of $n_s = 2.0\times 10^{10}$ cm$^{-2}$ for the n-GaAs sample, and 2.0 $\times 10^5$ cm$^2$V$^{-1}$s$^{-1}$  at
$n_s=2.0 \times 10^{11}$ cm$^{-2}$ for the p-GaAs sample. The $T$-dependence of $\sigma_{xx}$ shows a transition from insulating to
metallic behavior at $n_s = 5.17 \times 10^9$ cm$^{-2}$ (n-GaAs) and 4.50 $\times 10^{10}$ cm$^{-2}$ (p-GaAs). Firstly, we will
outline the procedures used to extract the interaction correction and $F^\sigma_0$ from the experimental measurements of
$\sigma_{xx}$ and $\rho_{xy}$ for the n-GaAs sample. After extracting the interaction correction and $F^\sigma_0$ for the p-GaAs
data, we will present a comparative analysis of $F^\sigma_0$ for the two samples, and discuss the ramifications of our measurements
with respect to the theory of Ref. \onlinecite{Zala}.

We begin by discussing the interaction correction to $\sigma_{xx}$ and $\rho_{xy}$ in n-GaAs because it is close to the ideal
system considered in Ref. \onlinecite{Zala} -- electrons in n-GaAs have spin 1/2, no valley degeneracy, and well defined effective
mass $m^*=0.067 m_e$. Furthermore additional corrections from phonon scattering and weak localization are negligible for the range
of experimental parameters that we have explored.

\begin{figure}[tbph]
\includegraphics[width=9cm]{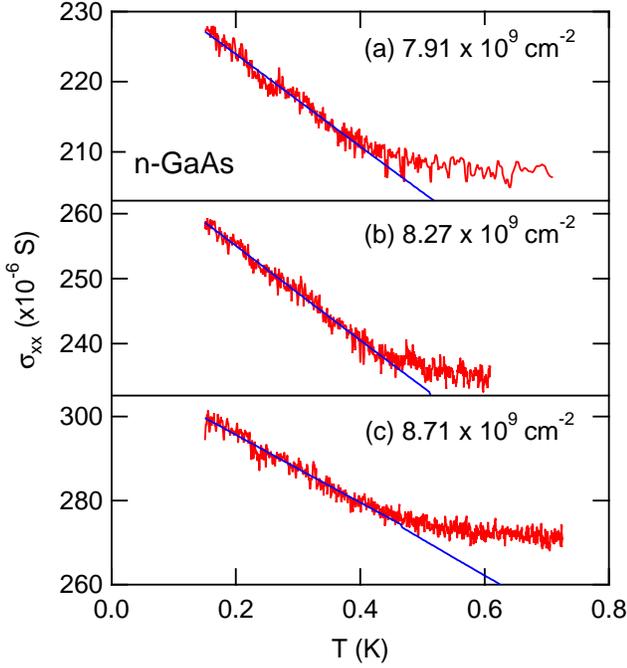}\caption{(a-c) Longitudinal conductivity $\sigma_{xx}$ vs temperature $T$ from the
n-GaAs sample at the indicated carrier densities. The solid lines are fits of Eqn. \ref{eqn1} to the data.} \label{Tdep_B0}
\end{figure}

Commencing with the longitudinal conductivity correction, the temperature dependence of $\sigma_{xx}$ at $B$=0 can be written as:

\begin{equation}\label{eqn1}
\sigma_{xx}(T) = \sigma_D + \delta\sigma_S(T) + \delta\sigma_T(T)
\end{equation}

\noindent where $\sigma_D$ is the Drude conductivity. The terms $\delta\sigma_S(T)$ and $\delta\sigma_T(T)$ are the singlet and
triplet channel interaction corrections given by Zala \emph{et al.} as \cite{Zala}:

\begin{eqnarray}\label{fit_dS}
\delta\sigma_S &=& \frac{e^2}{\pi h} \left[f(k_BT\tau)\frac{k_BT\tau}{h}-\ln\left(\frac{E_F}{T}\right)\right]\\\nonumber
\delta\sigma_T &=& \frac{e^2}{\pi
h}\left[g(k_BT\tau;F^\sigma_0)\frac{k_BT\tau}{h}-h(F^\sigma_0)\ln\left(\frac{E_F}{T}\right)\right]
\end{eqnarray}

\noindent where $f(x)$, $g(x;F^\sigma_0)$ and $h(x)$ are detailed in Ref. \onlinecite{Zala}. The procedure for extracting
$F^\sigma_0$ involves fitting Eqn. \ref{eqn1} to the graph of $\sigma_{xx}$ vs. $T$ in the degenerate limit $T\ll T_F$. The Fermi
energy $E_F$ and momentum scattering time $\tau$ are given by $E_F=\pi\hbar^2n_s/m^*$ and $\sigma_D=n_se^2\tau/m^*$ respectively
using measured values of $n_s$ and $\sigma_D$. This leaves $F^\sigma_0$ as the only fitting parameter in Eqn. \ref{eqn1}.

Figure \ref{Tdep_B0} (a-c) shows $\sigma_{xx}$ vs. $T$ for the n-GaAs sample at three values of $n_s$. At all carrier densities,
$\sigma_{xx}$ decreases linearly with increasing $T$ up to $\sim$0.5 K, above which $\sigma_{xx}$ levels off, and after reaching a
minimum value (not shown), increases again for higher $T$. This deviation from the low $T$ linear behavior has been observed
previously and is due to the 2D system becoming non-degenerate. \cite{DasSarma,ARH} Fits of Eqn. \ref{eqn1} to the experimental
data are shown as solid lines in Fig. \ref{Tdep_B0}. The fitting was limited to $T/T_F\leq 0.15$ where non-degenerate behavior is
not observed. The resulting values of $F^\sigma_0$ extracted from the fits, both with and without accounting for phonon scattering,
are found to be very similar (within 3$\%$)\cite{Karpus} and lie in the range $-0.4<F^\sigma_0<-0.36$; we will return to discuss
these extracted $F^\sigma_0$ values in more detail later.

We now turn to the correction to $\rho_{xy}$, which despite involving a longer procedure to extract, is significantly simpler than
the correction to $\sigma_{xx}$. This is because effects such as weak localization and phonon scattering have no impact on the
$T$-dependence of $\rho_{xy}$ and can be safely ignored. The Hall resistivity is composed of three terms

\begin{equation}\label{fit_Hall}
\rho_{xy}(T) = \rho_H^D + \delta\rho^\rho_{xy}(T) + \delta\rho^\sigma_{xy}(T)
\end{equation}

\noindent where $\rho_H^D= -B/n_se$ is the classical Hall resistivity, $\delta\rho^\rho_{xy}$ and $\delta\rho^\sigma_{xy}$ are the
singlet and triplet interaction correction terms given by: \cite{Zala_Hall}

\begin{eqnarray}
\frac{\delta\rho^\rho_{xy}}{\rho_H^D} &=& \frac{e^2}{\pi^2\hbar\sigma_D}
\left[\frac{11\pi}{192}\frac{\hbar}{k_BT\tau}+\ln\left(1+\frac{11\pi} {192}\frac{\hbar}{k_BT\tau}\right)\right]\\\nonumber
\frac{\delta\rho^\sigma_{xy}}{\rho_H^D} &=& \frac{3e^2}{\pi^2\hbar\sigma_D}
f(F^\sigma_0)\ln\left(1+\frac{11\pi}{192}\frac{\hbar}{k_BT\tau}\right)
\end{eqnarray}

\noindent where $f(x)$ is detailed in Ref. \onlinecite{Zala_Hall}.

The procedure for extracting $F^\sigma_0$ involves three steps: (i) The Hall slopes $d\rho_{xy}(T)/dB$ are obtained from the
$\rho_{xy}(T)$ traces as the gradient of linear fits over the range --20 mT $<B<$ 20 mT. (ii) In the high temperature limit both
$\delta\rho^\rho_{xy}$ and $\delta\rho^\sigma_{xy}\rightarrow$ 0 so that $\rho_{xy}(T)=\rho_H^D$, allowing $\rho_H^D$ to be
determined. Here, we take $\rho_H^D$ as $\rho_{xy}$ at 600 mK, the highest measurement temperature. (iii) $F^\sigma_0$ is then
obtained by fitting Eqn. \ref{fit_Hall} to a plot of $d\rho_{xy}(T)/dB$ vs. $k_BT\tau/\hbar$, with $\tau$ determined as specified
earlier and $F^\sigma_0$ as the only fit parameter.

\begin{figure}[tbph]
\includegraphics[width=9cm]{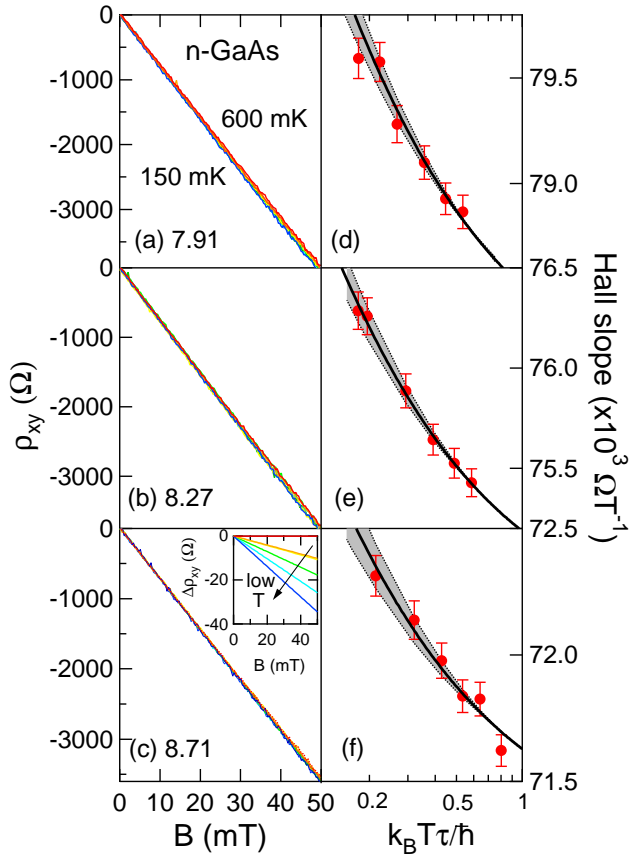} \caption{Left panel (a-c) shows the Hall resistivity $\rho_{xy}$ at the indicated carrier
densities (in $\times 10^9$ cm$^{-2}$) for the n-GaAs sample at temperatures ranging from 150 mK to 600 mK. Inset shows the
difference between the straight line fits at $T$ and 600 mK, $\Delta\rho_{xy}=\rho_{xy}(T)-\rho_{xy}(600mK)$, the arrow indicates
decreasing $T$. Values of the Hall slope are presented in the right panel (d-f) along with the fit. The shaded region indicates the
error in the fitting.} \label{Hallslope}
\end{figure}

Figure \ref{Hallslope} (a-c) shows plots of $\rho_{xy}(T)$ vs. $B$ at several temperatures in the range 150 mK $<T<$ 600 mK, for
$n_s$ values corresponding to those in Fig. \ref{Tdep_B0}. In each case the slope of $\rho_{xy}(T)$ vs. $B$ decreases by $\sim
1.5\%$ as the temperature is raised from 150 mK to 600 mK. This temperature-induced slope change is more clearly shown in Fig.
\ref{Hallslope} (c) (inset) where we plot the difference between the straight line fits to the Hall resistivity at $T$ and 600 mK,
$\Delta\rho_{xy}(T)=\rho_{xy}(T)-\rho_{xy}(600mK)$ vs. $B$ as a function of $T$. We have confirmed that this slope change is not
due to variations in $n_s$ by tracking the Shubnikov-de Haas (SdH) oscillations as a function of $T$ -- we find that the SdH minima
occur at constant $B$ to within 0.1 $\%$ as $T$ is varied.

Figure \ref{Hallslope} (d-f) shows the extracted Hall slope $d\rho_{xy}(T)/dB$ as a function of $k_BT\tau/\hbar$.  Since it is
difficult to reliably cool 2D systems to temperatures below 100 mK \cite{Prus}, we only present experimental data where we have
confirmed that the electron temperature matches the lattice temperature using Arrhenius plots of the SdH minima.

Fits of Eqn. \ref{fit_Hall} to the $d\rho_{xy}(T)/dB$ vs. $k_BT\tau/\hbar$ data are shown as solid lines in Fig. \ref{Hallslope}
(d-f), the shaded region indicates one standard deviation of error for the fit. Before we discuss the values of $F^\sigma_0$
extracted from the n-GaAs data, we will first repeat this analysis to extract interaction corrections for the p-GaAs sample, which
also shows metallic behavior at $B$=0. Commencing with obtaining $F^\sigma_0$ from the $\sigma_{xx}$ correction, weak localization
effects are not negligible and hence we need to subtract the weak localization correction from the $B$=0 $\sigma_{xx}$ data before
fitting with Eqn. \ref{Tdep_B0} to extract values for $F^\sigma_0$. The weak localization correction to the conductivity was
obtained by fitting the low field magnetoconductivity using the theory of Dmitriev \emph{et al.}, \cite{Dmitriev} which is valid
over a wider range of $B$ than the simpler theory of Hikami \emph{et al.} \cite{Hikami,SMc}

\begin{figure}[tbph]
\includegraphics[width=9cm]{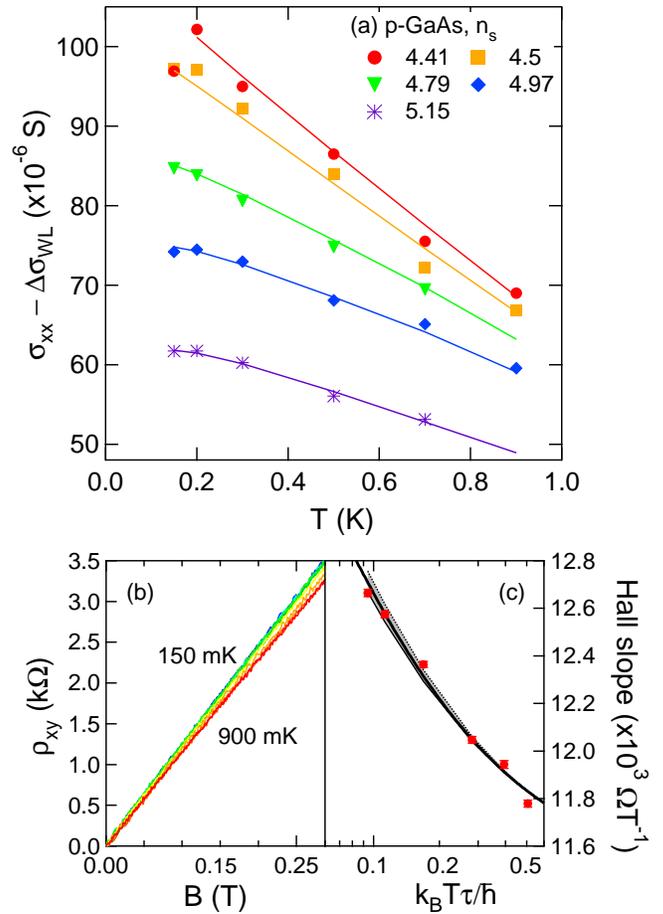}\caption{(a) Longitudinal conductivity $\sigma_{xx}$ of the p-GaAs sample after
subtracting the weak localization correction $\Delta\sigma_{WL}$ at the indicated carrier densities (in $\times 10^{10}$
cm$^{-2}$). (b) The Hall resistivity $\rho_{xy}$ at $n_s = 4.97 \times 10^{10}$ cm$^{-2}$ for different temperatures from 150 mK to
900 mK, and (c) the corresponding Hall slope along with the fit to Eqn. \ref{fit_Hall}. The shaded region indicates the error in
the fitting.} \label{A1433}
\end{figure}

The resulting values of $\sigma_{xx}-\Delta\sigma_{WL}$ for several different carrier densities are presented in Fig.
\ref{A1433}(a) along with the fits of Eqn. \ref{Tdep_B0}. At each $n_s$, the theory fits the experimental data well. The phonon
contribution is again small compared to $\delta\sigma_{xx}(T)$ and fitting with or without the phonon term gives similar values of
$F^\sigma_0$ (within 5$\%$);\cite{Karpus} this will be discussed later.

To obtain an independent measurement of $F^\sigma_0$ in the p-GaAs sample, we again study the $T$-dependent corrections to
$\rho_{xy}$. Fig. \ref{A1433}(b) presents a typical $\rho_{xy}$ trace at several different temperatures, showing a decrease in the
slope of $\rho_{xy}$ as $T$ is increased from 150 mK to 900 mK. Again we have confirmed that changes in the Hall slope are not due
to changes in carrier density. \cite{Simmons,MYS} The Hall slope $d\rho_{xy}(T)/dB$ vs. $k_BT\tau/\hbar$ is presented in Fig.
\ref{A1433}(c) along with a fit of Eqn. \ref{fit_Hall}, which describes the experimental data reasonably well. The fit quality for
the p-GaAs data is poorer than that of the n-GaAs and is likely due to the fact that p-GaAs is a more complex experimental system
(e.g. $m^*$ not well known, holes are spin 3/2 particles).

\begin{figure}[tbph]
\includegraphics[width=9cm]{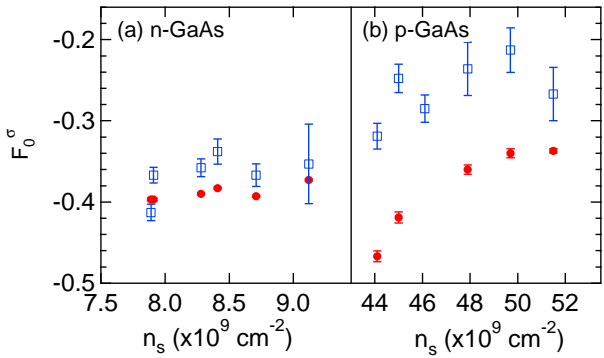} \caption{Values of the Fermi
liquid parameter $F^\sigma_0$ extracted from $\sigma_{xx}(T)$ at $B$=0 (solid) and $\rho_{xy}(T)$ (open) as a function of $n_s$ for
the (a) n-GaAs and (b) p-GaAs samples.} \label{F0}
\end{figure}

Fig. \ref{F0} contains the key result of our study. Firstly in Fig. \ref{F0} (a) we plot the values of $F^\sigma_0$ extracted from
$\sigma_{xx}$ (solid circles) and $\rho_{xy}$ (open squares) as a function of $n_s$ for the n-GaAs sample. The error bars
correspond to the fit errors discussed earlier (for the $\sigma_{xx}$ data in Fig. \ref{F0} (a), these are too small to show). As
the carrier density is increased, $F^\sigma_0$ increases in agreement with both theoretical expectations \cite{Zala} and previous
studies. \cite{Proskuryakov,Pudalov,Coleridge,Vitkalov} Most significantly at all carrier densities, the $F^\sigma_0$ values
extracted using the two different methods agree to within 9$\%$. The excellent agreement can only occur if the corrections to
$\sigma_{xx}$ and $\rho_{xy}$ originate from the same mechanism, thereby supporting the theory of Zala \emph{et al.} that these
corrections both derive from electron-electron screening of scattering processes.

Secondly, we plot the data for $F^\sigma_0$ extracted from $\sigma_{xx}$ (solid circles) and $\rho_{xy}$ (open squares) as a
function of $n_s$ in Fig. \ref{F0} (b) for the p-GaAs sample. A number of features stand out in this data. For both methods of
extracting $F^\sigma_0$ we see qualitatively similar behavior -- in general $F^\sigma_0$ decreases with decreasing $n_s$ -- to that
observed in the n-GaAs data. However, in contrast to the n-GaAs data, there is a marked quantitative discrepancy between the
$F^\sigma_0$ values extracted from the two methods, with the values differing by 30 $\%$. This discrepancy suggests that additional
corrections may be involved in the p-GaAs data, which is not entirely unexpected. Although the interaction effects are easier to
measure due to the significantly larger $r_s$ in p-GaAs, the holes are spin 3/2 particles and the effective mass is not well known
(we have taken it as $m^*=0.3m_e$ here), making comparisons between experimental data and the theory of Zala \emph{et al.}
\cite{Zala} a more complicated prospect. However, despite the quantitative discrepancy, the qualitative trends in the p-GaAs data
still support Ref. \onlinecite{Zala} and suggest that the anomalous $B$=0 metallic behavior is due to electron-electron or
hole-hole interactions, in agreement with previous studies in p-GaAs, \cite{Proskuryakov,Noh} n-Si, \cite{Pudalov} p-SiGe,
\cite{Coleridge} and n-GaAs. \cite{Li}

In summary, we present a study of the interaction correction in both 2D GaAs electron and hole systems using the theory of Zala
\emph{et al.} \cite{Zala,Zala_Hall} We find that independent measurements of the interaction correction from the $B$=0
$T$-dependence of $\sigma_{xx}$ and the low-$B$ $\rho_{xy}$ are in excellent agreement for the n-GaAs system, and are
quantitatively consistent to within 30$\%$, and in good qualitative agreement in p-GaAs. This supports the theory of Zala \emph{et
al.}\cite{Zala} in explaining the anomalous $B$=0 metallic behavior as due to electron-electron interactions screening the impurity
potential.

We are grateful to B. N. Narozhny for helpful discussions. This work was funded by the ARC. CEY acknowledges support from the UNSW
IPRS scheme.

\end{document}